# Origin of planar Hall effect in type-II Weyl semimetal MoTe$_2$

D. D. Liang,[1,2] Y. J. Wang,[1] W. L. Zhen,[1] J. Yang,[1] S. R. Weng,[1] X. Yan,[1] Y. Y. Han,[1] W. Tong,[1] L. Pi,[1,2] W. K. Zhu,[1,*] and C. J. Zhang[1,3,†]

[1]*Anhui Province Key Laboratory of Condensed Matter Physics at Extreme Conditions, High Magnetic Field Laboratory, Chinese Academy of Sciences, Hefei 230031, China*

[2]*Hefei National Laboratory for Physical Sciences at Microscale, University of Science and Technology of China, Hefei 230026, China*

[3]*Institute of Physical Science and Information Technology, Anhui University, Hefei 230601, China*

Besides the negative longitudinal magnetoresistance (MR), planar Hall effect (PHE) is a newly emerging experimental tool to test the chiral anomaly or nontrivial Berry curvature in Weyl semimetals (WSMs). However, the origins of PHE in various systems are not fully distinguished and understood. Here we perform a systematic study on the PHE and anisotropic MR (AMR) of $T_d$-MoTe2, a type-II WSM. Although the PHE and AMR curves can be well fitted by the theoretical formulas, we demonstrate that the anisotropic resistivity arises from the orbital MR (OMR), instead of the negative MR as expected in the chiral anomaly effect. In contrast, the absence of negative MR indicates that the large OMR dominates over the chiral anomaly effect. This explains why it is difficult to measure negative MR in type-II WSMs. We argue that the measured PHE can be related with the chiral anomaly only when the negative MR is simultaneously observed.

# I. INTRODUCTION

Weyl fermions in condensed matter systems represent the linearly dispersing low-energy excitations that obey a two-component Dirac equation [1]. Different from the type-I Weyl semimetals (WSMs, materials hosting Weyl fermions) which have standard Weyl points with a point-like Fermi surface, the type-II WSMs possess tilted Weyl points, arising at the contact of electron and hole pockets [2]. The orthorhombic phase ($T_d$) of layered transition metal dichalcogenides (TMDs) WTe$_2$ and MoTe$_2$ are theoretically predicted as the potential candidates for the type-II WSMs [2,3]. Signatures of Fermi arcs, the surface state of WSMs, have been indeed observed in the angle-resolved photoemission spectroscopy (ARPES) measurements for WTe$_2$ [4,5] and MoTe$_2$ [6,7].

In transport experiments, WSMs are usually featured for the negative longitudinal magnetoresistance (LMR) induced by the chiral anomaly [8,9], which refers to the non-conservation of chiral charge around the Weyl nodes when the applied electric and magnetic fields are nonorthogonal ($E \cdot B \neq 0$). The experimental measurement of negative LMR is very critical, and especially for type-II WSMs, the negative LMR can only be observed along specific crystalline directions and in samples with special components (see Refs. [10,11] for the negative LMR in WTe$_2$). Within our knowledge scope, so far, the negative LMR has not been reported for MoTe$_2$. Additionally, the measurement of negative LMR often suffers from several extrinsic effects, such as the current jetting effect [12-14]. Nevertheless, the negative LMR is still the most important transport method to investigate and identify WSMs.

Besides the negative LMR, planar Hall effect (PHE) is another transport evidence for the chiral anomaly or nontrivial Berry curvature in WSMs [15,16]. PHE is a well-known phenomenon in ferromagnetic metals and semiconductors [17,18], originating from the interplay of magnetic order and spin-orbit interactions. In WSMs, the chiral anomaly induced PHE ($\rho_{yx}$) and related anisotropic magnetoresistance (AMR, $\rho_{xx}$) are expressed as [15]

$$\rho_{yx} = -\Delta\rho^{chiral}\sin\theta\cos\theta, \quad (1)$$

$$\rho_{xx} = \rho_\perp - \Delta\rho^{chiral}\cos^2\theta, \qquad (2)$$

where $\Delta\rho^{chiral} = \rho_\perp - \rho_\parallel$ is the chiral anomaly induced anisotropic resistivity, $\rho_\perp$ and $\rho_\parallel$ are the resistivity corresponding to the magnetic field perpendicular to and along the direction of the current flow (*I*), respectively. Distinct features are revealed by Eq. (1). Namely, the angular dependence of $\rho_{yx}$ has a period of $\pi$, and its maximums appear at $\pm 45°$ and $\pm 135°$. This is different from the angular dependence measured in a usual Hall effect (period in $2\pi$) [19]. PHE has been experimentally observed in several topological semimetals [14,19-23]. However, the origins of PHE in these systems and other materials [24,25] may not be fully distinguished and understood. For the PHE caused purely by the chiral anomaly, the $\rho_\perp$ should be a constant, i.e., $\rho_0$, the $\rho_{xx}$ at zero magnetic field. Hence, the increase of $\Delta\rho^{chiral}$ with *B* only arises from the reduction of $\rho_\parallel$, i.e., the negative LMR. That is, the chiral anomaly induced PHE is just the angular dependence of negative LMR. The AMR is the anisotropy of the chiral transport. PHE cannot provide additional evidence for chiral anomaly, if the negative LMR is not observed. Actually, the experimentally measured PHE always has other contributions, such as orbital magnetoresistance (OMR).

In this paper, we perform a systematic study on the PHE and AMR of $T_d$-MoTe$_2$, in which the negative LMR induced by chiral anomaly is not observed. Although the PHE and AMR curves can be well fitted by the above equations, we demonstrate that the anisotropic resistivity $\Delta\rho$ arises from the orbital MR that increases the $\rho_\perp$ more quickly, instead of the reduced $\rho_\parallel$ as expected in the chiral anomaly effect. In contrast, the positive LMR indicates that the large orbital MR dominates over the chiral anomaly effect. This explains why it is difficult to measure negative LMR in type-II WSMs. We argue that the measured PHE can be related with the chiral anomaly only when the negative LMR is simultaneously observed.

## II. EXPERIMENTAL METHODS

$T_d$-MoTe$_2$ single crystals were grown by a self-flux method (Te). Powders of Mo (Alfa Aesar, 99.9%) and Te (Alfa Aesar, 99.99%) were ground and placed into a quartz ampoule, then heated up to 1373 K. The ampoule was cooled down to 1223 K at a rate of 2 K/h. The excess Te flux was removed by centrifugation. Finally, long flake-like crystals were obtained [inset of Fig. 1(a)]. The crystal structure and phase purity were checked by single crystal X-ray diffraction (XRD) on a Rigaku-TTR3 X-ray diffractometer using Cu Kα radiation. The single crystals are exfoliated into a thin plate to perform the transport measurements, with dimensions of 4×0.5×0.2 mm$^3$. All the electrical measurements were taken on a Quantum Design PPMS. Standard four-probe technique was used to measure longitudinal resistivity and Hall contacts were located on the transverse sides. The magnetic field was applied and rotated within the sample plane.

### III. RESULTS AND DISCUSSION
#### A. Sample characterizations

Figure 1(a) presents the single crystal XRD pattern taken at room temperature, which is consistent with the *1T'* phase of MoTe$_2$ [26]. Only the (00*l*) peaks are detected, suggesting that the naturally cleaved surface is the *ab* plane. Such a distorted octahedral phase will transit to an orthorhombic *T$_d$* phase at low temperature. This transition shows a signature in the resistivity measurement as a function of temperature, where a kink is observed at about 250 K [Fig. 1(b)] [27]. A relatively large residual resistivity ratio, i.e., RRR $= \frac{\rho_{300K}}{\rho_{2K}} = 156$, is further indicative of the high quality of the sample.

In order to obtain the mobility, the (regular) Hall effect and magnetoresistance (MR) are measured on the *ab* plane, with the magnetic field applied along the *c* axis [inset of Fig. 1(c)]. As shown in Fig. 1(c), the linear magnetic field dependence of $\rho_{xy}$ suggests a normal nature of the Hall effect. Although the carriers consist of both holes and electrons and a nearly compensated situation is supposed [26], the Hall effect is dominated by the electron-type carriers, as evidenced by the negative slope. Figure 1(d) shows the MR curve taken at 2 K, as well as the power law fit. The resultant component of n=1.87 agrees with the nearly compensated situation (n=2 for perfect compensation). Assuming $n_e = n_h$ in a two-band model [28], the MR ratio $\frac{\rho_{xx}(B) - \rho_{xx}(0)}{\rho_{xx}(0)}$ is equal to $\mu_e \mu_h B^2$, where $n_e$ ($n_h$) and $\mu_e$ ($\mu_h$) are the carrier density and mobility for electrons (holes), respectively. Using this formula to fit the MR curve yields a geometric-mean mobility $\mu_m = \sqrt{\mu_e \mu_h}$=5300 cm$^2$/Vs [inset of Fig. 1(d)]. This value is not as ultrahigh as in Cd$_3$As$_2$ (~10$^6$ cm$^2$/Vs) [29] or TaAs (~10$^5$ cm$^2$/Vs) [30], but almost that of Na$_3$Bi and GdPtBi (3000 and 2000 cm$^2$/Vs at 2 K, respectively) [21]. Such a relatively low mobility will lead to a large onset field $B_c$ (inversely proportional to mobility) of current jetting effect. The $B_c$ of Na$_3$Bi and GdPtBi can reach as high as 30 T [21]. Therefore, the current jetting effect is unlikely to be observed in the following PHE measurements for MoTe$_2$ (not more than 14 T). This is different from the case in TaP which has an ultrahigh mobility and enhanced current jetting effect [14].

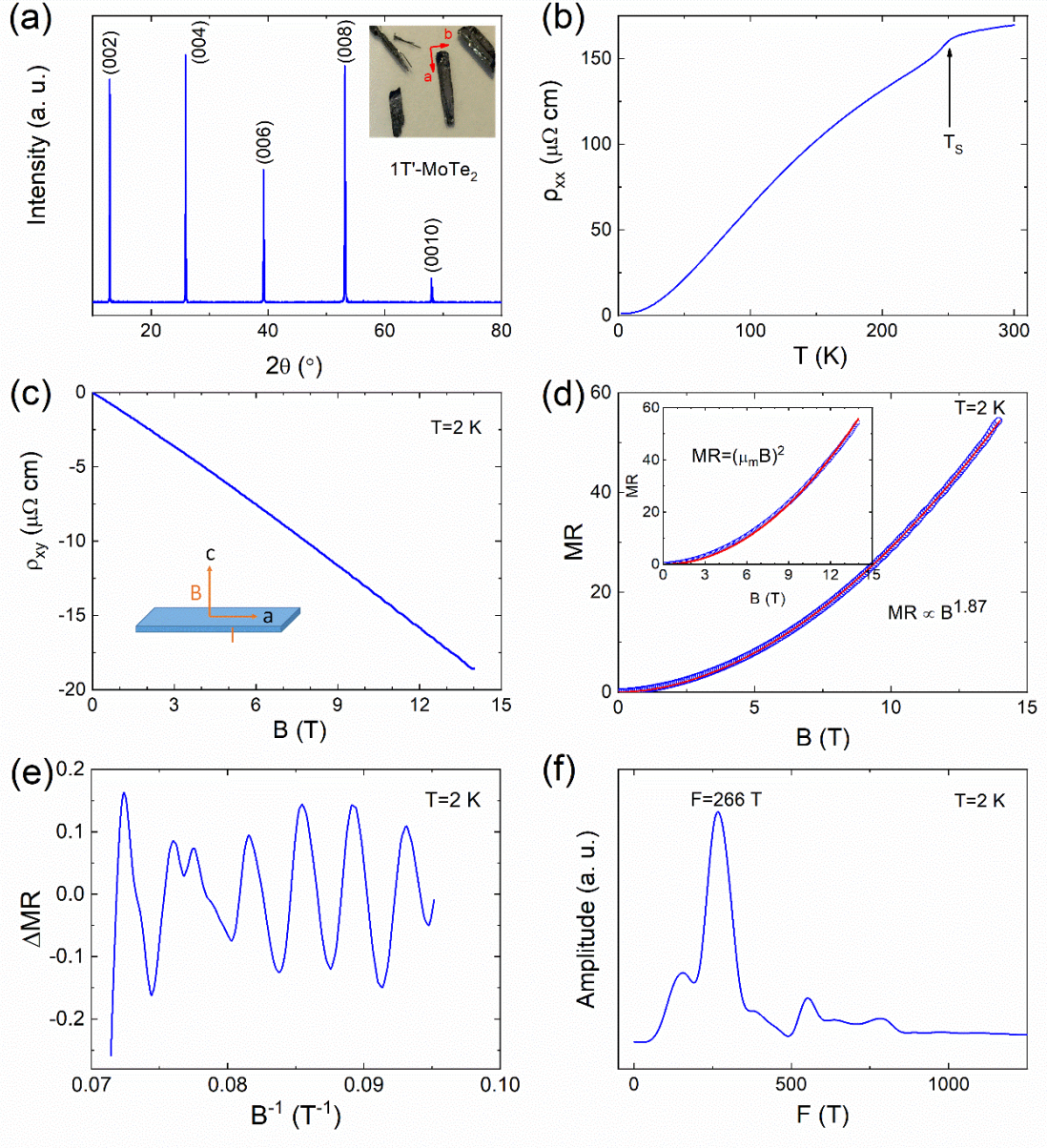

FIG. 1. Characterizations of MoTe$_2$ single crystal. (a) Single crystal XRD pattern taken at room temperature, consistent with the *1T'* phase of MoTe$_2$. Inset: image of as-grown single crystals. Red arrows indicate the crystalline axes. (b) Longitudinal resistivity $\rho_{xx}$ as a function of temperature taken at zero magnetic field, showing a first order phase transition at ~250 K. (c) Regular Hall resistivity $\rho_{xy}$ taken at 2 K in a magnetic field up to 14 T. Inset: the configuration of magnetic field and current flow for Hall effect and MR measurements. (d) MR ratio taken at 2 K in a magnetic field up to 14 T. Red solid curve represents the power law fit using the formula MR = $aB^n$. Inset: MR and the fitting curve using MR = $(\mu_m B)^2$. (e) Oscillating component extracted from the MR by subtracting the non-oscillatory background, plotted against $B^{-1}$. (f) Fast Fourier transformation spectra of the oscillation in (e).

The sample is further characterized by the Shubnikov-de Haas (SdH) oscillation analysis. The SdH oscillation can be observed when the magnetic field is above 10 T, and becomes even more pronounced after subtracting the non-oscillatory background (plotted against 1/$B$) [Fig. 1(e)]. By performing fast Fourier transformation, the oscillation frequency (i.e., a predominant branch $F$=266 T) is retrieved [Fig. 1(f)], consistent with previous researches [31].

## B. Giant planar Hall effect

The measurement geometry of PHE and AMR is illustrated in Fig. 2(a). The standard four-probe technique is adopted to measure the longitudinal MR ($\rho_{xx}$) along the $a$ axis, together with two Hall contacts to measure the planar Hall resistivity ($\rho_{yx}$). Magnetic field is applied within the sample plane ($ab$ plane) and rotates around the $c$ axis, with an angle $\theta$ relative to the current direction. Before analyzing the PHE and AMR data, we consider the possible extrinsic effects introduced by two types of misalignment in actual experimental set-up. Type-I misalignment is that the magnetic field does not always perfectly lie in the sample plane. Namely, a small out-of-plane field component may exist (thus a small deviation angle $\delta$), which will induce a small regular Hall resistivity in the measured $\rho_{yx}$. One way to eliminate this term is to average the $\rho_{yx}$ data taken at $\theta$ and $\theta + \pi$. Type-II misalignment is the possible nonsymmetrical Hall contacts that will induce a small longitudinal MR in the measured Hall resistivity. One part of the additional longitudinal MR is caused by the in-plane field, which has a $\cos^2\theta$ dependence according to Eq. (2), and another part is caused by the out-of-plane field, with an approximately $\sin^2(\theta + \delta)$ dependence. We note that both terms are symmetrical for $\pm\theta$, which is distinctly different from the odd-function feature of the PHE curve in Eq. (1).

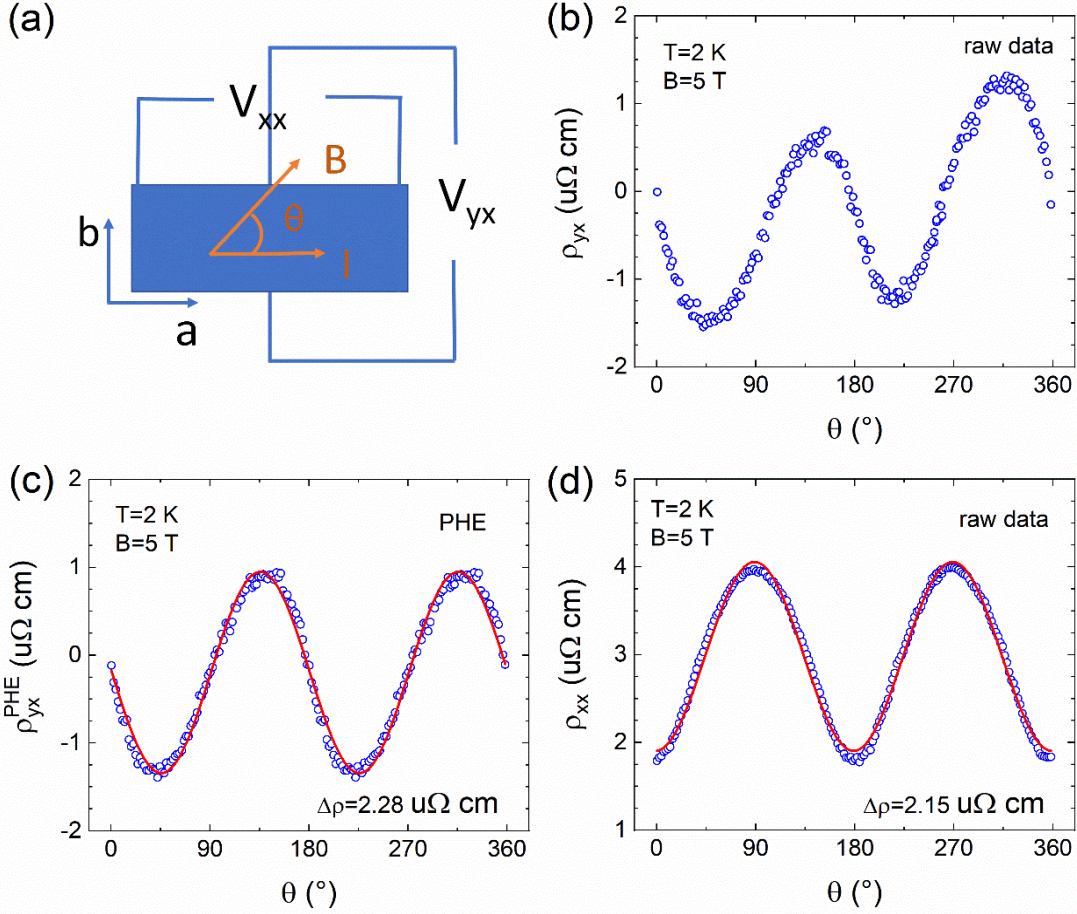

FIG. 2. (a) Schematic measurement geometry of planar Hall effect. Magnetic field is applied and rotated within the sample and current plane, with an angle $\theta$ relative to the current direction. (b) Raw data of the angle-dependent $\rho_{yx}$ taken at 2 K and 5 T. (c) The obtained planar Hall resistivity $\rho_{yx}^{PHE}$ after the averaging operation of $\theta$ and $\theta + \pi$. Red solid curve represents the fit to Eq. (1). (d) Raw data of the angle-dependent anisotropic MR $\rho_{xx}$ taken at 2 K and 5 T. Red solid curve represents the fit to Eq. (2).

As shown in Fig. 2(b), the raw data of $\rho_{yx}$ taken at 2 K and 5 T exhibits a slight slope, the footprint of regular Hall resistivity. According to the above analysis, the averaging operation of $\theta$ and $\theta + \pi$ is carried out to eliminate the regular Hall resistivity. The residual curve $\rho_{yx}^{PHE}$ shows an odd function feature [Fig. 2(c)], which thus suggests that the type-II misalignment is negligible in our measurements. The angular dependence of $\rho_{yx}^{PHE}$ has a period of $\pi$ and reaches its maximums at $\pm 45°$ and $\pm 135°$, both of which are consistent with the planar Hall effect. Using Eq. (1) to fit the experimental data results in an anisotropic resistivity $\Delta \rho = 2.28$ u$\Omega$ cm. This value is comparable with that of ZrTe$_5$ [23] and WTe$_2$ [22], but smaller than that of GdPtBi [19,21], Na$_3$Bi [21] and TaP [14]. Figure 2(d) presents the angular dependence of $\rho_{xx}$ taken at 2 K and 5 T. The type-I misalignment mentioned above also induces a normal MR component in the measured $\rho_{xx}$. However, this effect could be also neglected if we note that the resultant $\Delta \rho$ from the AMR curve (2.15 u$\Omega$ cm) is consistent with the $\Delta \rho$ obtained from the PHE curve. The slight difference between them should arise from the sample dimensions used in the resistivity calculations.

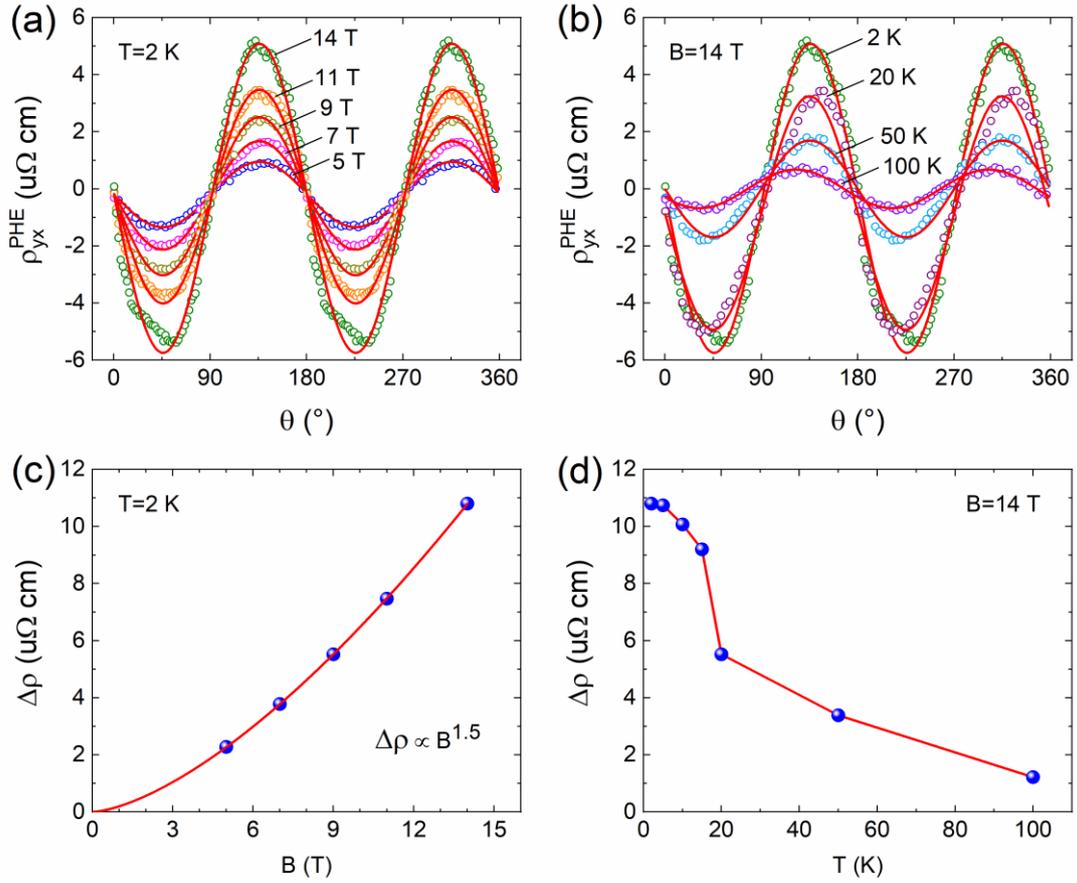

FIG. 3. Angular dependence of $\rho_{yx}^{PHE}$ (a) taken at 2 K and various magnetic fields and (b) taken at various temperatures and 14 T. Red solid curves represent the fits to Eq. (1). (c)

Magnetic field dependence of anisotropic resistivity $\Delta\rho$, obtained from the fits in (a). Solid curve is a power law fit. (d) $\Delta\rho$ as a function of temperature, obtained from the fits in (b).

The PHE measurements are further performed at different magnetic fields and temperatures, to test the magnetic field and temperature dependence of anisotropic resistivity $\Delta\rho$. Figure 3(a) shows the angular dependence of $\rho_{yx}$ taken at 2 K and various magnetic fields. We note that all the curves have a good sine-type line shape, confirming the absence of current jetting effect. The fittings to Eq. (1) give a series of $\Delta\rho$ for various $B$. As shown in Fig. 3(c), the magnetic field dependence of $\Delta\rho$ can be fitted to the power law with an exponent 1.5, which is divergent from the quadratic dependence that is expected for the chiral anomaly induced anisotropic resistivity. In addition to magnetic field, temperature is another factor to influence the $\Delta\rho$, via changing the mobility in $L_c/L_a$ [15], the scale of chiral charge transport. As a result of the enhanced thermal fluctuation, the mobility can be sharply reduced by the increasing temperature. Figure 3(b) presents the angular dependence of $\rho_{yx}$ taken at different temperatures and 14 T. The fitting results are plotted in Fig. 3(d), showing a remarkably decrease as tempeature increases.

### C. Anisotropic magnetoresistance and orbital magnetoresistance

The anisotropic resistivity $\Delta\rho$ can be also revealed by the angular dependence measurement of longitudinal resistivity $\rho_{xx}$, viz., the AMR. Figure 4(a) shows the angle-dependent $\rho_{xx}$ taken at 2 K and various magnetic fields. As magnetic field increases, the amplitude of $\rho_{xx}$ rises quickly. The fittings using Eq. (2) result in a list of $\Delta\rho$, which can be further fitted to the power law curve of $\Delta\rho \propto B^{1.4}$ [Fig. 4(c)]. We note that the values of $\Delta\rho$ and the fitting-resultant exponent are consistent with those in Fig. 3(c), again confirming the validity of our measurements and analyses.

To unveil the origin of $\Delta\rho$ that increases with $B$, the measured $\rho_\perp$ (i.e., $\rho_{xx}$ for $B \perp I$) and $\rho_\parallel$ (i.e., $\rho_{xx}$ for $B \parallel I$) are extracted from the $\rho_{xx}$ curves in Fig. 4(a). As seen in the inset of Fig. 4(c), the $\rho_\perp$ and $\rho_\parallel$ both increase as $B$ increases. However, their behaviors are different. With the increasing $B$, the $\rho_\perp$ rises rapidly ($\rho_\perp \sim B^{1.45}$), while the $\rho_\parallel$ increases in a moderate way. Some important information could be retrieved from the inset figure. First, the natural extensions of $\rho_\perp$ and $\rho_\parallel$ intersect at one point on the $y$ axis, i.e., $\rho_0$, the $\rho_{xx}$ at zero magnetic field. This value is consistent with the $\rho_{xx}$-$T$ measurement in Fig. 1(b). Second, the $\rho_\parallel$ shows a positive dependence on $B$, which means that the LMR is positive. This is contrary to the expectation of a negative LMR induced by the chiral anomaly. Although it is not concluded that the chiral anomaly does not exist in MoTe$_2$, it at least suggests that the chiral anomaly is dominated over by another factor that induces a positive LMR. Third, the main contribution of the increased $\Delta\rho$ ($=\rho_\perp - \rho_\parallel$) comes from the rapid increase of $\rho_\perp$, instead of the decrease of $\rho_\parallel$. We know that a nonzero $\Delta\rho$ can always produce a PHE curve, as is shown in Fig. 3(a). Therefore, the observation of PHE curves does not necessarily prove the existence of chiral anomaly. In this case, the PHE curves are associated with the excessively increased $\rho_\perp$.

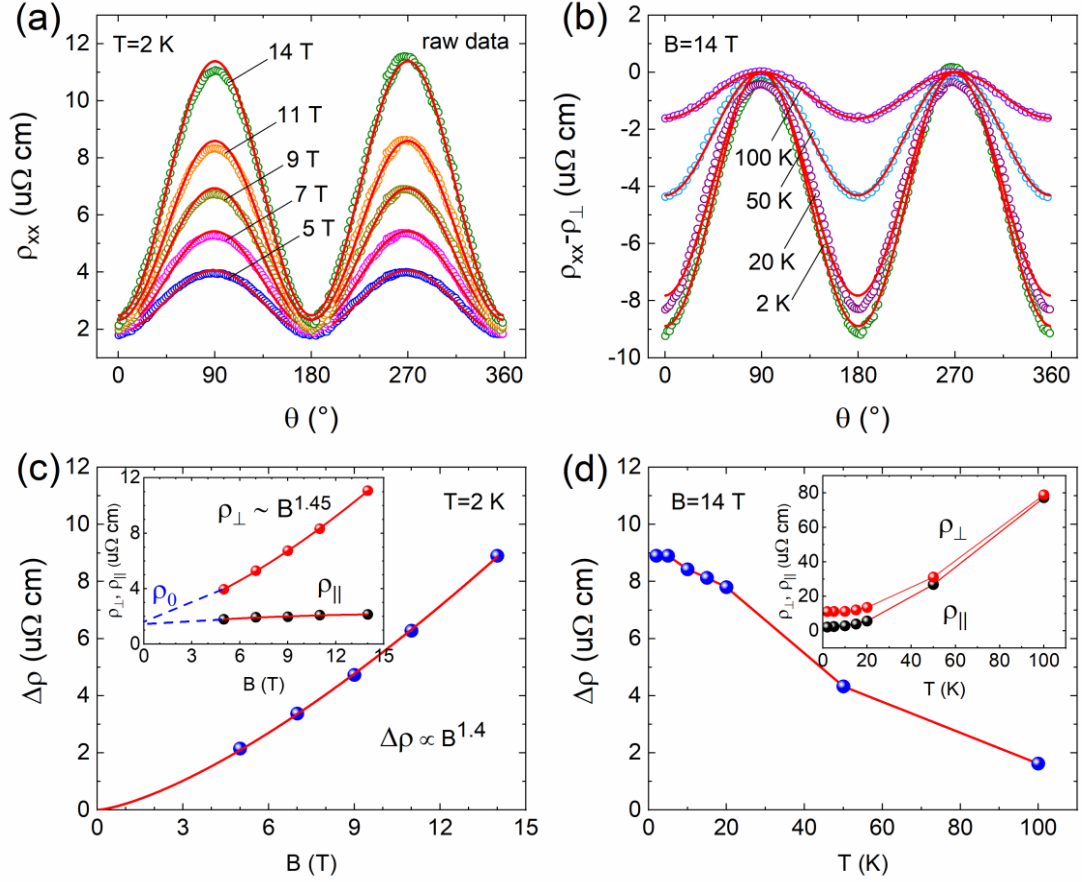

FIG. 4. (a) Raw data of the angle-dependent $\rho_{xx}$ taken at 2 K and different magnetic fields. (b) Angular dependence of $\rho_{xx} - \rho_\perp$ taken at different temperatures and 14 T. Red solid curves represent the fits to Eq. (2). (c) Magnetic field dependence of anisotropic resistivity $\Delta\rho$, obtained from the fits in (a). Solid curve is a power law fit. Inset: the $\rho_\perp$ and $\rho_\parallel$ extracted from the experimental data in (a). Solid curves represent the power law fit for $\rho_\perp$ and the 2nd order polynomial fit for $\rho_\parallel$, respectively. Blue dashed lines are the natural extensions which intersect at one point on the $y$ axis, i.e., $\rho_0$. (d) $\Delta\rho$ as a function of temperature, obtained from the fits in (b). Inset: the $\rho_\perp$ and $\rho_\parallel$ extracted from the experimental data in (b).

Since the magnetic field dependence of $\rho_\perp$ and $\rho_\parallel$ is far away from the quadratic dependence (n=1.87 in this case, see Fig. 1(d)), their increase with *B* is unlikely attributed to the addition of a normal MR due to the type-I misalignment. Taking account of the asymmetric Fermi surface of MoTe$_2$, we believe that the increase is arising from the orbital MR [32]. As suggested in Refs. [19,33], the orbital MR can significantly enhance the anisotropic conductivity $\Delta\sigma$ ($=\sigma_\parallel - \sigma_\perp$), and the magnetic field dependence of $\Delta\sigma$ is completely the same with the chiral anomaly induced PHE. Therefore, it is difficult to distinguish the origin of the $\Delta\rho$ measured in PHE, unless the magnetic field dependence of $\rho_\perp$ and $\rho_\parallel$ is also provided. As shown in the inset of Fig. 4(c), the orbital MR is highly anisotropic, increasing quickly for $\rho_\perp$ but slowly for $\rho_\parallel$. Hence, the existence of a large orbital MR is the origin of PHE. Actually in a pure PHE from chiral anomaly, the $\rho_\perp$ should be a constant, i.e., $\rho_0$. The increase of $\Delta\rho^{chiral}$ is mainly contributed by the decrease of $\rho_\parallel$, namely, the negative LMR. For most cases, the negative LMR is very small, because the $\rho_0$ of semimetals is usually small. For the present case, assuming that the chiral anomaly indeed exists and reduces the $\rho_\parallel$ to zero at a certain magnetic field, the largest contribution to $\Delta\rho$ is $\rho_0$ (~1.5 uΩ cm). However, this value is not sufficient to account for the large increase of $\Delta\rho$ (~8.9 uΩ cm at 14 T). The signature of chiral anomaly is fully covered by the orbital MR. This may explain why it is difficult to measure negative LMR in MoTe$_2$. For type-II WSMs, the orbital MR far exceeds the chiral anomaly induced negative LMR, which prevents the observation of chiral anomaly. Here, we argue that the measured PHE can be related with the chiral anomaly only when the negative LMR is simultaneously observed, corresponding to a situation of small orbital MR. Otherwise, the measured PHE is mainly from the anisotropic resistivity caused by orbital MR.

Figure 4(b) shows the temperature dependence of $\rho_{xx} - \rho_\perp$ taken at various temperatures and 14 T. Consistent with the results in 3(d), the $\Delta\rho$ is greatly reduced when temperature increases [Fig. 4(d)]. The $\rho_\perp$ and $\rho_\parallel$ are presented in the inset of Fig. 4(d), both increasing with $T$, as in Fig. 1(b). However, the difference of them becomes smaller at a high $T$. That is, the enhanced thermal fluctuation reduces the resistivity anisotropy, even in the presence of an external magnetic field. As discussed above, the reduced $\Delta\rho$ does not indicate the suppression of chiral anomaly effect, as the underlying mechanism is dominated by the orbital MR.

## IV. CONCLUSIONS

In summary, the angular dependence of PHE and AMR is studied in detail for the type-II WSM $T_d$-MoTe$_2$. We demonstrate that the anisotropic resistivity $\Delta\rho$ arises from the orbital MR that increases the $\rho_\perp$ more quickly, instead of the reduced $\rho_\parallel$ as expected in the chiral anomaly effect. In contrast, the positive LMR indicates that the large orbital MR dominates over the chiral anomaly effect. This explains why it is difficult to measure negative LMR in type-II WSMs. We argue that the measured PHE can be related with the chiral anomaly only when the negative LMR is simultaneously observed.

## ACKNOWLEDGMENTS

This work was supported by the National Key R&D Program of China (Grant Nos. 2016YFA0300404, 2017YFA0403600 and 2017YFA0403502), the National Natural Science Foundation of China (Grant Nos. U1532267, 11674327, 11874363, 51603207, 11574288 and U1732273). D.D.L. and Y.J.W. contributed equally to this work.